\documentclass[showpacs,preprintnumbers,amsmath,amssymb, nofootinbib]{revtex4}
\newcommand{\be}{\begin{equation}}
\newcommand{\en}{\end{equation}}
\newcommand{\bea}{\begin{eqnarray}}
\newcommand{\ena}{\end{eqnarray}}

\begin{document}
\title{ Bare and effective fluid description in brane world cosmology}
\author{Norman Cruz$^{1}$, Samuel Lepe$^{2}$, Francisco Pe\~na$^{3}$ and Joel Saavedra$^{2}$}
\address{$^{1}${\small Departamento de
F\'\i sica, Facultad de Ciencia, Universidad de Santiago, Casilla
307, Santiago, Chile,}}
\address{$^{2}${\small Instituto de F\'{\i}sica, Pontificia
Universidad Cat\'olica de Valpara\'{\i}so, Casilla 4950,
Valpara\'{\i}so, Chile,}}
\address{ $^{3}${\small Departamento de
Ciencias F\'\i sicas, Facultad de Ingenier\'\i a, Ciencias y
Administraci\'on, Universidad de La Frontera, Avda. Francisco
Salazar 01145, Casilla 54-D Temuco, Chile.} }
\date{\today}

\begin{abstract}
An effective fluid description, for a brane world model in five
dimensions, is discussed for both signs of the brane tension. We
found several cosmological scenarios where the effective equation
differs widely from the bare equation of state.  For universes
with negative brane tension, with a bare fluid satisfying the
strong energy condition, the effective fluid can cross the barrier
$\omega _{eff} =-1$.

\end{abstract}


\maketitle

\section{\label{sec:level1} Introduction}

Consideration of  dimensionality of the the universe greater than
four has had a long history passed from the original idea of
Kaluza-Klein \cite{kaluza} to modern ideas of string theory
\cite{Font:2005td, Chan:2000ms}. In particular the Randall-Sundrum
scenario has acquired a great attention in the last decade
\cite{Randall:1999ee, Randall:1999vf}. From the cosmological point
of view, brane world offers a novel approach to our understanding
of the evolution of the universe. The most spectacular consequence
of this scenario is the modification of the Friedmann equation. In
those models, for instance in five dimensions, matter is confined
to a four dimensional brane, while gravity can be propagated in
the bulk, i.e., gravity is the only field that can feel the extra
dimension. From the perspective of string theory ~\cite{witten},
brane world cosmology has been a big challenge for modern
cosmology . For a comprehensible review on BW cosmology see
Ref.~\cite{lecturer}. For example, consequences of a chaotic
inflationary universe scenario in a BW model was
described~\cite{maartens}, where it was found that the slow-roll
approximation is enhanced by the modification of the Friedman
equation. In the Einstein approach the DE appears as an exotic
fluid in the energy-momentum tensor, so the field equations
\cite{Gabadadze:2007dv} becomes
\begin{equation}
G_{\mu \nu }=8\pi G_{N}\left( T_{\mu \nu }^{matter}+T_{\mu \nu
}^{DE}\right) ,  \label{intro1}
\end{equation}
where $G_{\mu \nu }$ is the $4D$ Einstein tensor, $T_{\mu \nu
}^{matter}$ is the stress tensor for matter and $T_{\mu \nu
}^{DE}$ is the stress tensor for dark energy, a new exotic
component with a negative pressure. This description can be
modified in such way that this exotic component arises naturally
from modifications in the geometric sector of the field equations,
\begin{equation}
G_{\mu \nu }-\mathcal{K}_{\mu \nu }=8\pi G_{N}T_{\mu \nu
}^{matter}, \label{intro 2}
\end{equation}
where $\mathcal{K}_{\mu \nu }$ denotes a tensor that arise from
the extrinsic curvature, due to the embedding of our brane
universe in the $5D$ bulk (for a review see Ref. \cite{rm} and
references therein). Using projection techniques
\cite{Shiromizu:1999wj} it was found the effective Friedmann
equation onto the brane, which can be written as follows,
\begin{equation}
3H^{2}=8\pi G_{N}f(\rho ),  \label{intro3}
\end{equation}
where the function $f(\rho)$ encoded all geometric modification of
the field equations. In this framework exotic matter it is not
necessary to explain the late acceleration of the universe and the
coincidence problem. In the context of this framework, the
evolution of universes filled with a perfect fluid has been
investigated in many works, see for example Refs. \cite{lecturer,
maartens, Zhang:2007yu, Shtanov:2002mb}.

The aim of this article is discuss the behavior of the brane world
models in five dimensions for a positive and negative brane
tension, in terms of an bare and effective fluid description.

The plan of the paper is as follows: In Sec. II we specify the
effective four dimensional cosmological equations from
Randall-Sundrum model. In Sec III we give the principal equations
corresponding an effective fluid description for the cosmological
evolution. We show the behavior of the found scenarios for a
positive and negative brane tension. In Sec. IV we discuss the
cosmological scenarios in terms of the bare and effective
description.


\section{\label{sec:level2} Randall-Sundrum
Cosmological Scenario}


For an homogeneous and isotropic flat 4-brane, described by the
Friedmann-Lemaitre-Robertson-Walker (FLRW) metric, the field
equations are
\begin{equation}
3H^{2} =\rho \left( 1\pm \frac{\rho }{2\lambda }\right) ,
\label{eq.1}
\end{equation}
where the positive and negative sign are related to positive  and
negative brane tension. The matter content sector satisfy the
conservation equation given by
\begin{equation}
\dot{\rho }+3H\left( \rho +p \right) =0.  \label{eq.2}
\end{equation}
Deriving Eq. (\ref{eq.1}) with respect to the cosmological time
and using Eq. (\ref{eq.2}) we obtain
\begin{equation}\label{acceleration1}
\frac{\ddot{a}}{a}=-\frac{1}{6}(\rho+3p)\left[1\pm\frac{\rho}{\lambda}\right]
\mp\frac{\rho^2}{6\lambda}.
\end{equation}
In what follows we consider a barotropic the equation of state
$p=\omega\rho$ for the bare fluid.


\section{Effective fluid description on the brane}


Since we are interested in an effective description of the
modified Friedmann equations obtained from the Randall-Sundrum
models, we define the dimensionless variable, $x=\rho /2\lambda $.
This allow us to rewrite Eq. (\ref{eq.1}) in the standard form
\begin{equation}
3H^{2} =\rho _{eff},  \label{eq.3}
\end{equation}
where the effective density is given by
\begin{equation} \rho _{eff} =2\lambda
x\left( 1\pm x\right). \label{effectiverho}
\end{equation}
The quadratic term in the $x$ variable describes brane world
correction on the cosmological equations. Taking the derivative of
Eq. (\ref{eq.3}) with respect to cosmological time we obtain
 \be \dot{H} =-\frac{1}{2}\left[
\rho _{eff}+p_{eff}\right], \label{hpunto} \en where the effective
pressure is given by
\begin{eqnarray}
p_{eff} &=&  p \left( 1\pm 2 x\right) \pm 2\lambda x^2 .
\label{eq.6}
\end{eqnarray}
In this sense, the effective state equation is given by
\begin{equation}
\omega _{eff}=\frac{p_{eff}}{\rho _{eff}}=\frac{1}{1\pm x}\left[
\omega\left( 1\pm 2x\right)\pm x\right],  \label{eq.9}
\end{equation}

Meanwhile, the effective fluid satisfy the usual equation of
conservation
\begin{equation}\label{eq.conseff}
\dot{\rho}_{eff}+3H\left( \rho_{eff} +p_{eff} \right) =0.
\end{equation}
The acceleration in terms of the variable $x$ is given by
\begin{equation}\label{eq.conseff}
\frac{\ddot{a}}{a}=-\frac{\lambda}{3}\left[(1+3\omega)x(1\pm2x)\pm2x^2\right].
\end{equation}

This effective scenario allows us to describe the Randall-Sundrum
cosmology in the standard way and we can fix or constraint the
cosmological parameters. In the following we are discussing this
effective scenarios and we particularize in the sign of the brane
tension.


\subsection{SCENARIOS FOR A POSITIVE BRANE TENSION}


\textbf{ $\bullet$ \,Accelerated universes with a inflexion point
}

\vspace{0.5cm}

In the case of a brane with positive tension the modified
Friedmann becomes,
\begin{eqnarray}\label{F00tensionpositiva}
3H^2=2\lambda x (1+x).
\end{eqnarray}
The acceleration is given in this case by
\begin{eqnarray}\label{aceleraciontensionpositiva}
\frac{\ddot{a}}{a}=-\frac{\lambda}{3}
  \left[(1+3\omega)x(1+2x)+2x^2\right].
\end{eqnarray}
If we search for an scenarios where exist a inflexion point
between phases of acceleration and deceleration, it is necessary
to impose the existence of an inflexion point, $x_t$, where
$\ddot{a}(x_t)=0$. This point is given by
$x_{t}=-\frac{1}{2}\left(\frac{1+3\omega}{2+3\omega}\right)$.
Since we must have $x_{t}>0$, the equation of state of the bare
matter is constrained to the
range$-\frac{2}{3}<\omega<-\frac{1}{3}$. Note that this implies
that the inflexion point exist the strong energy condition is
violated, i.e.,  $1+3\omega <0$.

The acceleration given in Eq. (\ref{aceleraciontensionpositiva})
can be rewritten in the form
\begin{eqnarray}\label{aceleraciontensionpositivatt}
\frac{\ddot{a}}{a}=-\frac{2\lambda}{3}(2+3\omega)(x-x_t)x.
\end{eqnarray}
Using Eq. (\ref{eq.2}) with a constant $\omega$ we obtain a
decreasing energy density of the bare fluid as the universe
expand, then in early times the universe begin enters in a
decelerated phase, reach the point of inflexion and then begin to
accelerate.  This scenarios resembles the future evolution of the
universe at late times, as we observe today.  In order to discuss
the how behaves effective cosmological fluid we rewrite the
effective equation of state, given by Eq. (\ref{eq.9}), in terms
of the inflexion point, yielding
\begin{equation}\label{1+3wefectivotensionpositiva}
1+3\omega_{eff}=2\left(\frac{2+3\omega}{1+x}\right)\left(x-x_{t}\right).
\end{equation}
Note that independent of the $\omega$, at the inflexion point,
$x_{t}$, the effective fluids is $\omega_{eff}=-1/3$.

\vspace{0.5cm}

\textbf{ $\bullet$ \,Accelerated universes without a inflexion
point}

\vspace{0.5cm}

Other possible cosmological evolutions not include a inflexion
point, which occurs whenever $\omega> -1/3$, or $\omega< -2/3$,.
The first condition is obviously satisfied by pressureless bare
matter or bare matter with positive pressure, i.e., $\omega\geq
0$, driven an universe which evolves always with $\ddot{a}<0$.
Eq.(\ref{eq.9}) indicates that the effective fluid also has a
positive pressure. In general, the effective fluid has higher
pressure that the bare one.  Since at early times we can assume
$x\sim 1$, we obtain that $\omega_{eff}\sim (3\omega +1)/2$. If,
for example, the bare matter is radiation,  the effective fluid is
stiff matter at early times. At late times we see that
$\omega_{eff}\sim \omega $.  Therefore, in general the effective
fluid is diluting with the cosmic evolution. An universe with
radiation as bare matter expand with deceleration and the
effective matter evolves from stiff matter to radiation.   An
scenario derived in \cite{Banks}, for an holographic cosmology,
leads to a FLRW universe with equation of state $p = \rho$.

For $\omega \leq -2/3$, which includes phantom matter, evolves
always with an positive acceleration.  From Eq.(\ref{eq.9}) is
clear to see that if $\omega =-1$, corresponding a cosmological
constant in the bare matter, we obtain also $\omega_{eff}=-1$, for
the effective fluid.  Note that even for bare phantom matter, i.
e., $\omega <-1$, the effective fluid never becomes a phantom
fluid.

It is interesting to mention that if we have a phantom bare
matter, the universe could begin with a low density, which is
increasing as the cosmic times evolves.


\subsection{SCENARIOS FOR A NEGATIVE BRANE TENSION}


\vspace{0.5cm}

\textbf{ $\bullet$ \,Accelerated universes with a inflexion point}

\vspace{0.5cm}

For this case the Friedmann equation is it is interesting to note
that, this kind of modified Friedmann equation can be derived also
from effective loop quantum cosmology \cite{lqc}
\begin{equation}
3H^{2} = 2\lambda x\left( 1-x\right),
\end{equation}
where we have now the constraint $0<x<1$. Let us consider first
scenarios where exist a inflexion point in the acceleration. Then
in this cases the acceleration is given by
\begin{equation}
\frac{\ddot{a}}{a }\left( x\right) =-\frac{2\lambda}{3} \left(
2+3\omega \right) x\left( x_{t}-x\right) ,
\end{equation}
where  the inflexion point, $x_{t}$, is given by
\begin{equation}\label{xt}
x_{t}=\frac{1}{2}\left( \frac{1+3\omega }{2+3\omega }\right).
\end{equation}
The condition $x_{t}>0$ implies the constraint $\omega > -1/3$ or
$\omega < -2/3$.  If we consider the first constraint then the
strong energy condition $\rho +3p=\left( 1+3\omega \right) \rho
>0$ is satisfied by the bare matter. In order to discuss the effective
behavior we shall to use the effective equation of state given, in
this case, by
\begin{equation}\label{weff}
1+3\omega _{eff}=\frac{2}{1-x}\left ( 2+3\omega \right ) \left(
x_{t}-x\right).
\end{equation}
If $\omega > -1/3$ then $2+3\omega>0$, so for $x<x_{t}$ the
universe is decelerating , with $\omega _{eff} > -1/3$ and for
$x>x_{t}$ is accelerating with $\omega _{eff} < -1/3$. According
to this, whereas $\omega$ satisfied the strong energy condition,
the effective fluid can be described by normal matter,
quintessence or phantom matter. It is interesting to pointed out
here an important different between the standard description and
the effective description. It is well-known in the framework of
standard cosmology that for universes filled with one fluid,
satisfying  the strong energy condition, the evolution shall be
driven by just one accelerated phase . In the effective
description for a brane with negative tension, where the bare
matter satisfy, the strong energy condition we found the presence
of both phases: accelerated and decelerated.

Since the variable $x$ is decreasing with the cosmic time, at the
early times $x\leq x_{t}$, which indicates that initially we have
an accelerated phase and then an decelerated phase when $0\leq
x\leq x_{t}$. Note that this scenario can occurs with a bare fluid
with positive pressure, in other words, with normal matter.

It is direct to verify that
\begin{equation}
\frac{\ddot{a}(x)}{\ddot{a}(1/2)}=2 (1+3\omega)x
\left(\frac{x}{x_{t}} -1\right),
\end{equation}
and from the inequalities
\begin{equation}
\ddot{a}(0<x<x_{t}) < \ddot{a}(x_{t}<x<1/2) < \ddot{a}(1/2)<
\ddot{a}(1/2<x<1),
\end{equation}
where $\ddot{a}(0<x<x_{t})<0$ and all the accelerations in the
ranges indicates above are positives. Note that $ \ddot{a}/a (1/2)
= \lambda /6$ correspond to a the Sitter phase just at the point
$x=1/2$.  In brief, the above results indicates that the effective
matter behaves initially as normal matter following to
quintessence, de Sitter and phantom matter.


An interesting and novel content can be found if we relax the
strong energy condition and we consider that the condition
$1+3\omega <0$ is satisfied. From the condition $0<x_{t}<1$ we
shall investigate the the constraint upon the the equation of
state of the mater. If we rewrite Eq. (\ref{xt}) in the following
form
\begin{equation}
x_{t}=\frac{1}{2}\left( \frac{\left| 1+3\omega \right| }{\left|
1+3\omega \right| -1}\right),
\end{equation}
we obtain that  if $x_{t}>0$ then $| 1+3\omega | >1$ and the
constraint is $\omega <-2/3$. If $x_{t}<1$ then $| 1+3\omega |
>2$, which leads to the condition  $ \omega <-1$.  Since both
constraint must be satisfied we conclude that the matter confined
on the brane corresponds to a phantom matter. It is easy to check
for bare phantom matter, $ \frac{1}{2}<x_{t}<1$. In this scenario,
the bare phantom matter drives an accelerated phase at the
beginning, i.e., for $x<x_{t}$. After the inflexion point,
$x>x_{t}$, the universe end in a decelerated phase. A remarkable
point of this phantom scenario is that does not reach a future
singularity. This kind of phantom behavior was discussed in Ref.
\cite{Curbelo:2005dh} for an universe with arbitrary (non
gravitational) interaction between the components of the cosmic
fluid. It was found a wide region in the parameter space where the
solutions are free of the big rip singularity, suggesting that
phantom models without big rip singularity might be preferred by
nature.  In terms of the effective fluid, the universe is
accelerating , with $\omega _{eff} < -1/3$ and is decelerating
with $\omega _{eff} > -1/3$.

\vspace{0.5cm}

\textbf{ $\bullet$ \,Accelerated universes without a inflexion
point}

\vspace{0.5cm}

If the strong energy condition is violated, there are models which
not include an inflexion point whenever $-1\leq\omega\leq -1/3$.
This indicate that the  bare matter is a quintessence fluid.
Notice that that the particular case of $\omega=-1$ implies
$\omega_{eff}=-1$. In other words, if the bare matter is a
cosmological constant, the effective fluids also behaves like
cosmological constant during all the cosmic evolution driving an
accelerated expansion.

Eq.(\ref{eq.9}) indicates that the effective fluid behaves during
the cosmic evolution from $x\sim 1$,at early times, to $x\sim 0$
going from $\omega_{eff}\ll w$ to $\omega_{eff}\sim w$,
respectively. Obviously, in this scenario the universe is always
accelerated.


\section{\label{sec:level3}Discussion}

In this article we have studied brane world cosmology in term of
the effective fluid description on the brane.

We have found two types of scenarios for a positive and negative
brane tension: those that show an inflexion point in the
acceleration and those that present a continuous acceleration or
deceleration during the cosmic evolution. For universes with a
positive brane tension, the first scenario exist if the strong
energy condition is violated for the bare matter. In this case the
universe evolves initially with deceleration and then begins to
accelerate at the inflexion point $x_{t}$.  This scenario
resembles the observed evolution of the universe today from
supernova data. In the decelerated phase, the effective fluid has
the equation of state $\omega_{eff}>-1/3$. Note that this not
exclude that at early times the effective matter could be, for
example, dust.  In the accelerated phase, the effective fluid has
the equation of state $\omega_{eff}<-1/3$, which leads to the
possibility to have cosmological constant or phantom as effective
matter.

In the other scenario, for positive brane tension, there is no
inflexion point for the acceleration. For bare matter with $\omega
> -1/3$, which include bare matter with positive pressure, the
effective fluid has an equation of state with higher pressure that
bare one. For example, a universe filled with radiation as bare
matter expand with deceleration and the effective matter evolves
from stiff matter to radiation.  For bare matter with $\omega <
-1/3$, which include phantom matter, the universe evolves always
with a positive acceleration. Although, the bare matter could be
phantom matter, the effective matter never becomes a phantom
fluid.  In this case the early universe has low density.

For a negative brane tension, accelerated universes with an
inflexion point, occurs for the both cases of a bare matter
satisfying and violating the strong energy condition.  If $\omega
>-1/3$, the universe begins accelerating with $\omega _{eff} < -1/3$,
and then is decelerating with $\omega _{eff} > -1/3$. According to
this, whereas $\omega$ satisfied the strong energy condition, the
effective fluid can be described by normal matter, quintessence or
phantom matter. At the effective level, the equation of state
presents a crossing of the barrier $\omega _{eff} =-1$.

When the strong energy condition is violated the bare phantom
matter drives an accelerated phase at the beginning, and after the
inflexion point the universe end in a decelerated phase. A
remarkable point of this phantom scenario is that does not reach a
future singularity. As it was said before, in terms of the
effective fluid, the universe is accelerating , with $\omega
_{eff} < -1/3$ and is decelerating with $\omega _{eff} > -1/3$.

There are models which not include an inflexion point whenever
$-1\leq\omega\leq -1/3$, i.e., when the bare matter is a
quintessence fluid. Exist the particular case $\omega=-1$, since
the effective equation of state is then $\omega_{eff}=-1$. In
other words, if the bare matter is a cosmological constant, the
effective fluids also behaves like cosmological constant during
all the cosmic evolution driving an accelerated expansion. In
general, the effective fluid behaves during the cosmic evolution
going from $\omega_{eff}\ll w$, at early times, to
$\omega_{eff}\sim w$, lately.

Within the possible scenarios described above, a remarkable
evolution is found when the effective description is taken account
This correspond to a universe with negative brane tension, with a
bare fluid satisfying the strong energy condition, although the
effective fluid could be quintessence or phantom. The effective
fluid can cross the barrier $\omega _{eff} =-1$. Also it is of
interest the case of a universe with radiation as bare matter,
expanding decelerated and with the effective matter evolving from
stiff matter to radiation. This scenario have been discussed in
the framework of holographic cosmology \cite{Banks}.

\begin{acknowledgments}
This work was supported by COMISION NACIONAL DE CIENCIAS Y
TECNOLOGIA through FONDECYT \ Grant 11060515 (JS). This work was
also partially supported by PUCV Grant No. 123.785/2006 (JS),
DIUFRO DI08-0041 of Direcci\'on de Investigaci\'on y Desarrollo
Universidad de la Frontera (FP) and by DICYT 040831 CM,
Universidad de Santiago de Chile (NC). The authors SL, JS and NC
wish to thank Departamento de F\'{\i}sica de la Universidad de La
Frontera for its kind hospitality.
\end{acknowledgments}

\end{document}